\documentclass[%
reprint,
superscriptaddress,
showpacs,
 amsmath,amssymb,
 prl
]{revtex4-1}

\usepackage{graphicx}
\usepackage{dcolumn}
\usepackage{bm}
\usepackage{xcolor}

\begin{document}

\title{Large-scale photonic Ising machine by spatial light modulation}

\author{D. Pierangeli}
\email{Davide.Pierangeli@roma1.infn.it}
\affiliation{Dipartimento di Fisica, Universit\`{a} di Roma  ``La Sapienza'', 00185 Rome, Italy}
\affiliation{Institute for Complex System, National Research Council (ISC-CNR), 00185 Rome, Italy}

\author{G. Marcucci}
\affiliation{Dipartimento di Fisica, Universit\`{a} di Roma  ``La Sapienza'', 00185 Rome, Italy}
\affiliation{Institute for Complex System, National Research Council (ISC-CNR), 00185 Rome, Italy}

\author{C. Conti}
\affiliation{Dipartimento di Fisica, Universit\`{a} di Roma  ``La Sapienza'', 00185 Rome, Italy}
\affiliation{Institute for Complex System, National Research Council (ISC-CNR), 00185 Rome, Italy}

\begin{abstract}
Quantum and classical physics can be used for mathematical computations that are hard to tackle by conventional electronics. Very recently, optical Ising machines have been demonstrated for computing the minima of spin Hamiltonians, paving the way to new ultra-fast hardware for machine learning. However, the proposed systems are either tricky to scale or involve a limited number of spins. We design and experimentally demonstrate a large-scale optical Ising machine based on a simple setup with a spatial light modulator. By encoding the spin variables in a binary phase modulation of the field, we show that light propagation can be tailored to minimize an Ising Hamiltonian with spin couplings set by input amplitude modulation 
and a feedback scheme.
We realize configurations with thousands of spins that settle in the ground state in a low-temperature ferromagnetic-like phase with all-to-all and tunable pairwise interactions. Our results open the route to classical and quantum photonic Ising machines that exploit light spatial degrees of freedom for parallel processing of a vast number of spins with programmable couplings.
\end{abstract}

\maketitle
A large number of internal states characterizes complex systems from biology to social science. The fact that the number of these states grows exponentially with the system size hampers large-scale computational possibilities. Complex optimization problems involving these models are in many cases classified as NP-hard and cannot be tackled efficiently by standard computing architectures. A broad class of such computationally intractable problems maps to the search of the ground state of a classical system of interacting spins: the minimization of an Ising Hamiltonian with specific spin couplings \cite{Barahona1982, Lucas2014, Cuevas2016}.

Growing research interest is emerging towards physical and artificial systems that evolve according to an Ising Hamiltonian and enable to find the optimal combinatorial solution by the ground state observed in the experiment.
Quantum and classical Ising systems have been realized by trapped atoms \cite{Kim2010, Britton2011}, single photons \cite{Ma2011}, superconducting circuits \cite{Johnson2011}, electromechanical modes \cite{Mahboob2016}, nanomagnets \cite{Sutton2017} and polariton
condensates \cite{Berloff2017}. In optics, spin-glass dynamics have been observed in random lasers \cite{Ghofraniha2015, Tommasi2016}, multimodal cavities \cite{Basak2016, Moura2017} , coupled laser lattices \cite{Nixon2013}, beam filamentation \cite{Ettoumi2015} and nonlinear wave propagation in disordered media \cite{Pierangeli2017}. These photonic systems host thousands of optical spins, but the spin variables are not easy to access
and controlling their interaction is challenging. 

Novel photonic platforms with numerous and easily accessible spins are particularly relevant for computation. Optical computing machines offer high-speed and parallelization. Various authors reported coherent Ising machines based on time-multiplexed optical parametric oscillators finding approximate solutions to optimization problems with several nodes \cite{Marandi2014, McMahon2016, Inagaki2016, Inagaki2016_2, Takeda2017, King2018, Bello2019}.
Others proposed nanophotonic circuits to implement any small-scale spin systems directly on a programmable chip \cite{Shen2017, Harris2017, RoquesCarmes2018}. 
Matrix operations can also be performed by spatially shaped optical fields, without engineered wave-mixing devices \cite{Huisman2015, Fickler2017},
by exploiting randomly reflected waves \cite{Hougne2018} or disordered biological samples \cite{Pierangeli2018}. 
However, using spatial optical modulation to solve Ising spin dynamics has remained unexplored.

In this Letter, we propose and experimentally demonstrate the use of spatial light modulation for calculating the ground state of an Ising Hamiltonian. The phase matrix on a spatial light modulator (SLM) acts as a lattice of spins whose interaction is ruled by the constrained optical intensity in the far-field and can be programmed by input amplitude modulation.
Feedback from the detection plane allows the spatial phase distribution to evolve towards the minimum of the selected spin model. We find ferromagnetic-like ground-states in agreement with mean-field predictions. Our spatial Ising machine hosts thousands of parallelly-processed spins, and represents a scalable and efficient approach for photonic computing. 

\begin{figure}[t!]
\centering
\vspace*{-0.1cm}
\hspace*{-0.2cm} 
\includegraphics[width=1.04\columnwidth]{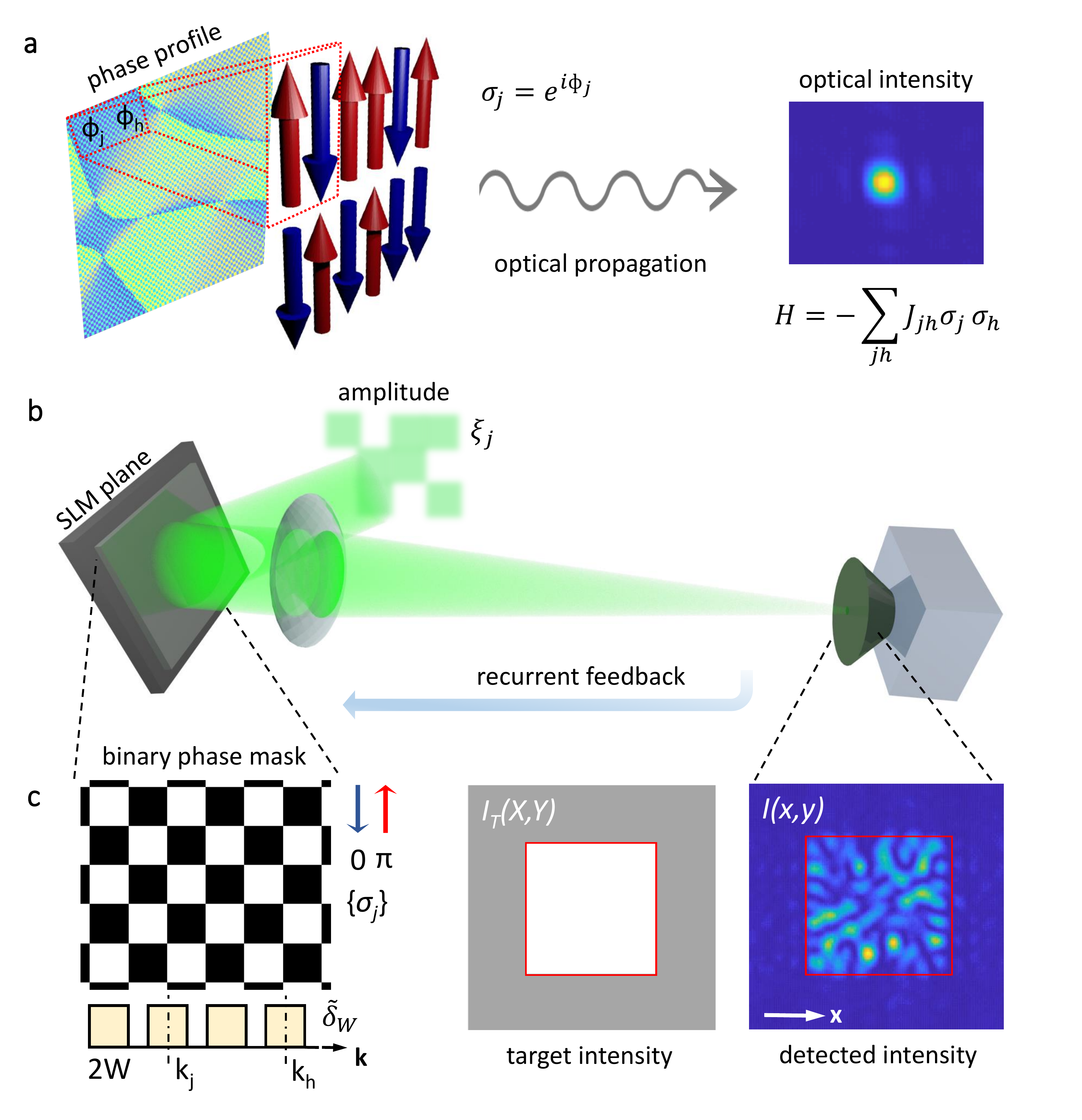} 
\vspace*{-0.5cm}
\caption{Ising machine by spatial light modulation. (a) The wave phase in different spatial points gives the spins evolving through optical propagation. (b) An amplitude-modulated laser beam is phase modulated by a reflective SLM and detected by a CCD camera in the far-field. (c) A discrete phase mask with binary values $\phi_j=0,\pi$ in the Fourier plane mimics Ising spins $\sigma_j=\pm 1$. 
Inset is an example of the detected intensity when the binary hologram is tailored to generate a squared intensity target $I_T$.
}
\vspace{-0.2cm}
\label{Figure1}
\end{figure}

We implement a spatial photonic Ising machine by using the phases in separated spatial points of the optical wavefront.
A binary phase modulated beam encodes binary spins with configurable interactions [Fig.~\ref{Figure1}(a)].
A spin variable $\sigma_j=\exp({\imath \phi_j})=\pm 1$ corresponds to a spatial point of the optical field with phase $\phi_j\in\{0,\pi\}$.
As illustrated in Fig.~\ref{Figure1}(b-c), an SLM acting as a reprogrammable matrix of pixels imprints binary phase values on the coherent wavefront. Setting the SLM in the Fourier space of the electric field $\tilde E(k)$, we have 
\begin{equation}
  \tilde E(k)= \sum_j \xi_j \sigma_j \tilde{\delta}_W(k-k_j),
  \label{Ek}
\end{equation}
where $\xi_j$ indicates the field amplitude incoming on each pixel.
The normalized rectangular function $\tilde{\delta}_W$ models the pixel of finite size $2W$ [Fig.~\ref{Figure1}(c)], so that $k_j= 2Wj$, with $j=1,...,n$. The resulting far-field intensity after free-space propagation is
\begin{equation}
I(x)=|E(x)|^2= \sum_{jh} \xi_j \xi_h \sigma_j \sigma_h \delta_W^2(x) e^{2\imath W(h-j)x},
\end{equation}
with $\delta_W(x)=\sin(W x)/ (W x)$ the inverse Fourier transform of $\tilde{\delta}_W(k)$.
Spin-spin interaction can be induced by acting on the intensity on the detection plane.
We constrain $I(x)$ by a measurement and feedback method to couple the phases on the SLM plane. 
Minimizing ${\Vert I_T(x) - I(x)\Vert}$ for an arbitrary target intensity image $I_T(x)$ thus corresponds to minimizing a Hamiltonian $H$.
After normalization $\int [I_T(x)]^2 \mathrm{d}x \simeq \int [I(x)]^2 \mathrm{d}x$, and the function $H$ takes the form of the Ising Hamiltonian
\begin{equation}
H= - \sum_{jh} J_{jh} \sigma_j \sigma_h
\label{eq:hamiltonian}
\end{equation}
with spin interactions given by 
\begin{equation}
J_{jh}= 2\xi_j \xi_h \int I_T(x) \delta_W^2(x) e^{2\imath W(h-j)x} \mathrm{d}x .
\label{eq:inter}
\end{equation}
When the effect of the SLM pixel size can be neglected, $\delta_W(x)\sim 1$, and the couplings reduce to 
\begin{equation}
J_{jh}= 2\pi \xi_j \xi_h \tilde{I}_T[2W(j-h)],
\label{eq:interft}
\end{equation}
which indicates that the interaction matrix is set by the input amplitude modulation along with the Fourier transform of the far-field target image.
The interaction passes from short- to long-range by changing the spatial profile of $I_T$.
In particular, in the case of a point-like target image, the spins are all-to-all interacting ($J_{jh}=\mathrm{const}$) for an input wave with constant amplitude. Using a programmable (quenched) amplitude mask on the input beam the couplings can be varied according to $J_{jh}\propto \xi_j \xi_h$, which allows to implement the entire class of spin-glass models, known as Mattis models \cite{Mattis1976, Nishimori2001},
where the pairwise interaction can be expressed as product of two independent variables. 
Fig.~\ref{Figure1}(c) shows the principle of operation of our Ising machine. 
A spin configuration $\{\sigma_{j}\}$ is generated upon an amplitude-modulated wavefront using binary phases on the SLM and the corresponding intensity distribution $I(x)$ is measured in the far-field. The detected image is compared with the target $I_T(x)$ and the information is feedback to the SLM plane. The system evolves towards minimization of the cost function  $f=||I_T(x) - I(x)||$ , which corresponds to the Ising ground state. 

The experimental optical machine follows the setting shown in Fig.~\ref{Figure1}(b).
Light from a continuous-wave laser source with wavelength $\lambda= 532$nm is expanded, eventually modulated in amplitude, and impinges on a  nematic liquid crystal reflective modulator (Holoeye LC-R 720, $1280\times768$ pixels, pixel pitch $20\times20 \mu$m) whose active area is selected by a rectangular aperture to host $N=L \times L$ spins (pixels). The SLM is set into a phase-modulation mode with less than 10\% residual intensity modulation by a combination of incident and analyzed polarizations. Phase-modulated light is spatially filtered ($3$mW power) and then focused by a lens (f$=500$mm) on a CCD camera. The intensity is measured on $M=18\times18$ spatial modes obtained grouping $16 \times 16$ camera pixels to average over speckles arising from spatial phase fluctuations in the far-field plane. 

\begin{figure*}[t!]
\centering
\vspace*{-0.2cm}
\hspace*{-0.1cm} 
\includegraphics[width=1.95\columnwidth]{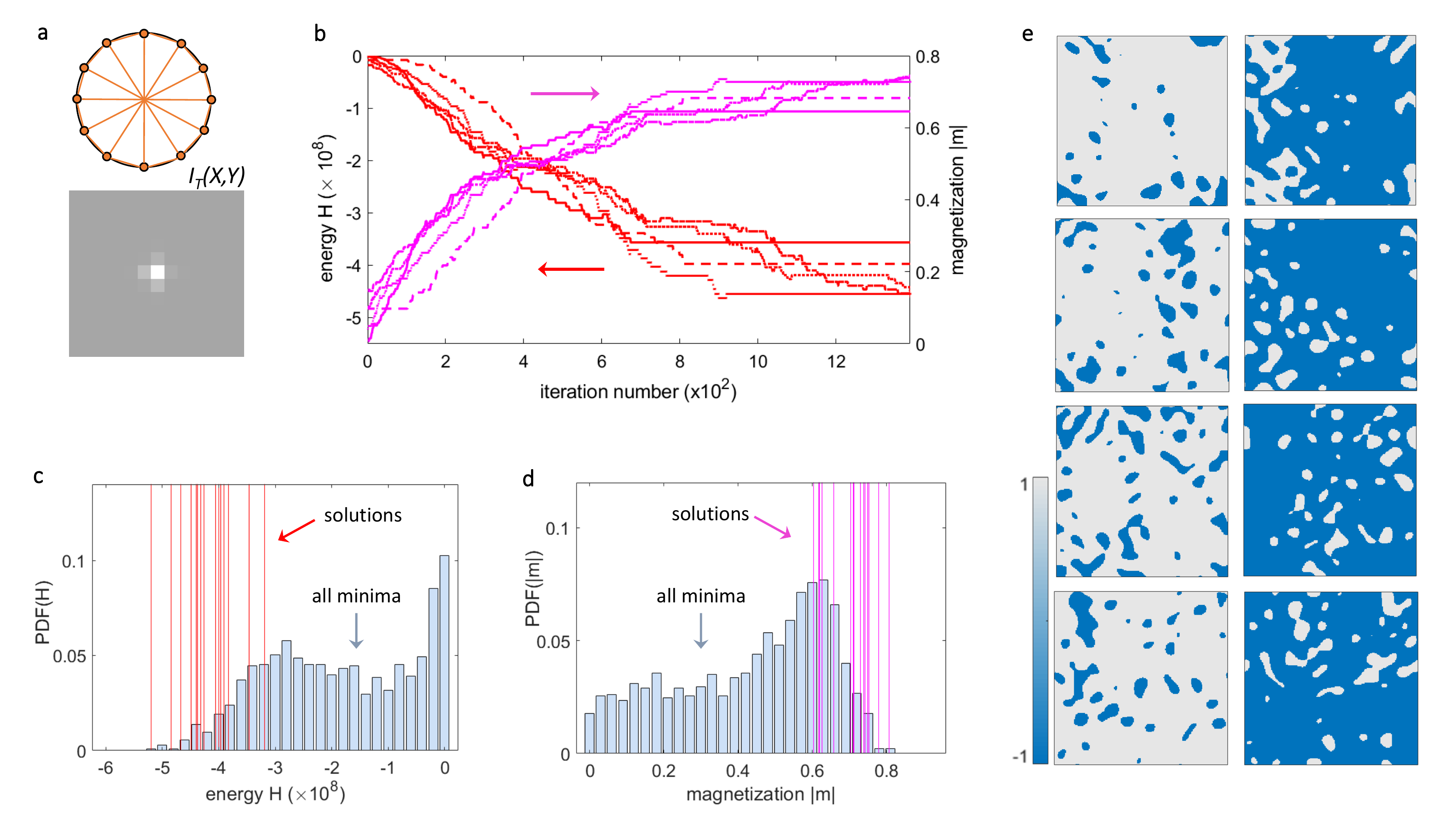} 
\vspace*{-0.2cm}
\caption{Optically solving the Ising Hamiltonian with all-to-all spin interactions. (a) An unweighted M\"obius-Ladder graph with fully-connected vertices (results refer to $N=4\times 10^4$) along with the target intensity $I_T$. (b) Measured evolution of $H$ ($J_{jh}=1$) and $|m|$ for different initial random spin matrices. (c-d) Observed probability distribution for the (c) energy and (d) magnetization of spin configurations satisfying the interaction constrain $I_T$; red and magenta lines indicate $H$ and $|m|$ of the identified ground-state solutions, respectively.
(e) A set of ground-state spin configurations: small-size ferromagnetic clusters with opposite magnetization are visible.}
\vspace{-0.1cm}
\label{Figure2}
\end{figure*}

We first demonstrate the spatial Ising machine for $N=4\times 10^4$ spins with all-to-all couplings ($J_{jh}=\mathrm{const}$),
which corresponds to a number of spin-spin connections orders of magnitude larger than those realized in time-multiplexed platforms
\cite{McMahon2016, Inagaki2016}.
In this case, $\xi_j=\xi_h=\xi_0$ and the target corresponds to intensity focused only in a single spatial mode, that is, a bright localized spot [Fig.~\ref{Figure2}(a)]. The binary phases on the SLM are initialized by a random distribution, which gives a weak and broad speckle pattern in the detection plane. By a Monte Carlo-like method, at each iteration we randomly flip a small cluster of spins and measure the corresponding far-field intensity, retaining the change only if its difference with the target image decreases \cite{Vellekop2008}.
Unlike other photonic Ising machines \cite{King2018}, no information about the target Hamiltonian is used to affect electronically the spin evolution.
To prevent trapping into local minima induced by the algorithm, we select clusters with a gradually increasing size.
To follow the system evolution we consider as physical observables the energy $H$ and the magnetization $m= \langle \sigma_j \rangle$ of each configuration. 
As shown in Fig.~\ref{Figure2}(b) for different realizations, we observe a monotonic growth of $|m|$, which saturates to a large
value after approximatively $10^3$ iterations. The Hamiltonian monotonically decreases toward a plateau, thus indicating the onset of a low-energy
ferromagnetic-like state. The actual temperature $T$ of these spin configurations is determined by the random phase fluctuations in the Fourier plane, which results from the intrinsic noise characterizing each operation in the experimental setup. Sources of noise come from the quantization on the CCD discrete modes of the detected intensity as well as from the imperfect spatial phase modulation \cite{Ritsch-Marte2007}.

To test the solution found by our machine, we use a different approach based on phase retrieval. The aim is to evaluate the energy probability distribution function (PDF) of all those $\{\sigma_j\}$ that satisfy the far-field constraint and compare with the low-energy solutions found by the machine. We use a quantized phase-retrieval (QPR) algorithm \cite{Wyrowski1990} to generate binary phase distributions from the target
image $I_T$ and measure the far-field intensity $I$. Among the many QPR states, which are associated with different phase patterns in the target plane, the solution of the machine is determined by minimizing the cost function $f$.
Fig.~\ref{Figure2}(c-d) show the results from $16$ set of measurements, each with $100$ phase-retrieved spin configurations.
The identified solutions populate the tail of the energy distribution [Fig.~\ref{Figure2}(c)] and have maximum magnetization [Fig.~\ref{Figure2}(d)]. This indicates that ground states of the Ising Hamiltonian are successfully found.
In particular, the machine gives with $87$\% probability the correct minimum solution, that is, a spin configuration lying in the 5\% of those with the lowest energy. This ground state probability quantifies the correspondence 
between the cost-function minima and spin states with lower energy, and is independent of the way the ground state has been found. 

\begin{figure}[t!]
\centering
\hspace*{-0.1cm}
\includegraphics[width=0.98\columnwidth]{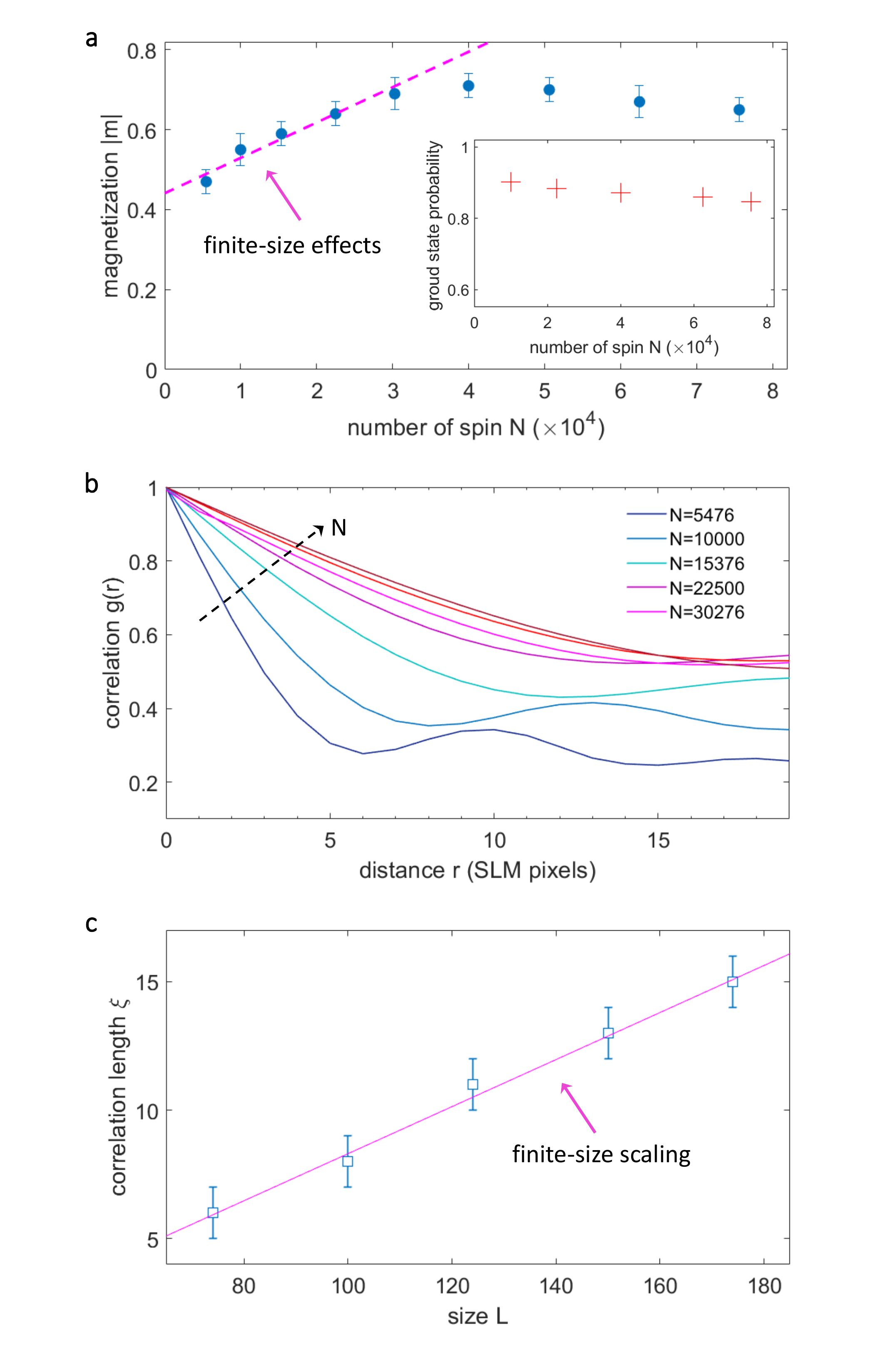} 
\vspace*{-0.2cm}
\caption{Scaling properties of the ferromagnetic ground state. (a) Observed magnetization varying the spin number. The inset shows the scaling of the machine performance. (b) Spatial spin autocorrelation functions (distance in pixel units) for different $N$. (c) Autocorrelation length as a function of the system size $L$ (dots) and linear fitting behavior (line).}
\vspace{+0.1cm}
\label{Figure3}
\end{figure}

To quantify the physical state resulting from the optical computation, we analyze the spin configurations. Fig.~\ref{Figure2}(e) shows the typical ground states retrieved by the optical machine. We observe ferromagnetic domains of various size embedded in a phase with opposite magnetization. Spin states with $m<0$ and $m>0$ appear with almost equal probability, as expected from spontaneous symmetry breaking in the absence of external magnetic fields. From the set of $\{\sigma_j\}$ we estimate the actual temperature according to the mean-field solution of Eq.~(\ref{eq:hamiltonian}), which describes the case with all-to-all interacting spins \cite{Parisi1988, Nishimori2001}.
Considering the equation of state $m=\tanh{ [(T_c/T)m] }$, from the observed mean magnetization we obtain $T/T_c=0.80 \pm 0.03$.
We also analyze the measured spin spatial autocorrelation according to $g(r)=\exp{(-r/\xi)}$, where the autocorrelation length $\xi$
estimates the mean domain size. In the mean-field approach $\xi$ diverges at the critical temperature as $\xi= R_*\left( 1-T/T_c \right)^{-\beta}$, where the critical exponent $\beta=1/2$ and $R_*$ is the minimum cluster length. The resulting temperature is $T/T_c=0.83 \pm 0.02$. Therefore, the observed ground states have magnetizations and domain configurations consistent with a mean-field Ising model at  fixed temperature.

One of the main features of our spatial photonic setting is the extremely large number of spins that can be simulated.
Varying the active area on the SLM (the transverse size of the spatially modulated laser beam), 
we investigate how the machine operation depends on the system size $L$. Fig.~\ref{Figure3}(a) shows the magnetization and the fidelity
(probability of finding the Ising ground state) of the observed ground state varying the number of spin from $N=74\times74$ to $N=274\times274$ and leaving unchanged their interaction. 
At variance with other photonic settings \cite{McMahon2016}, we find that the performance of our Ising machine does not sensibly depend on the number of spins [inset in  Fig.~\ref{Figure3}(a)]. For large sizes $N$, a minor decrease of the magnetization and fidelity is observed, 
and due to the lower spatial resolution in the detection plane.
At low spin number, we observe a linear decrease of $|m|$ as $N$ is reduced. We ascribe this behavior to finite-size effects. The observed spin autocorrelation function strongly varies with the number of spins, and a well-defined single decay only emerges at large $N$ [Fig.~\ref{Figure3}(b)]. 
For configurations with few spins, we find that the measured correlation length grows linearly with the configuration size [Fig.~\ref{Figure3}(c)], in agreement with finite-size scaling arguments, which predicts a mean-field behavior $\xi \propto L$ \cite{Landau1976}.
For large $L$ the size of ferroelectric domains becomes independent of the system scale. The photonic machine thus points out a fundamental phenomenon of spin models \cite{Binder1997}.

\begin{figure}[b!]
\centering
\vspace*{-0.1cm}
\hspace*{-0.2cm}
\includegraphics[width=1.03\columnwidth]{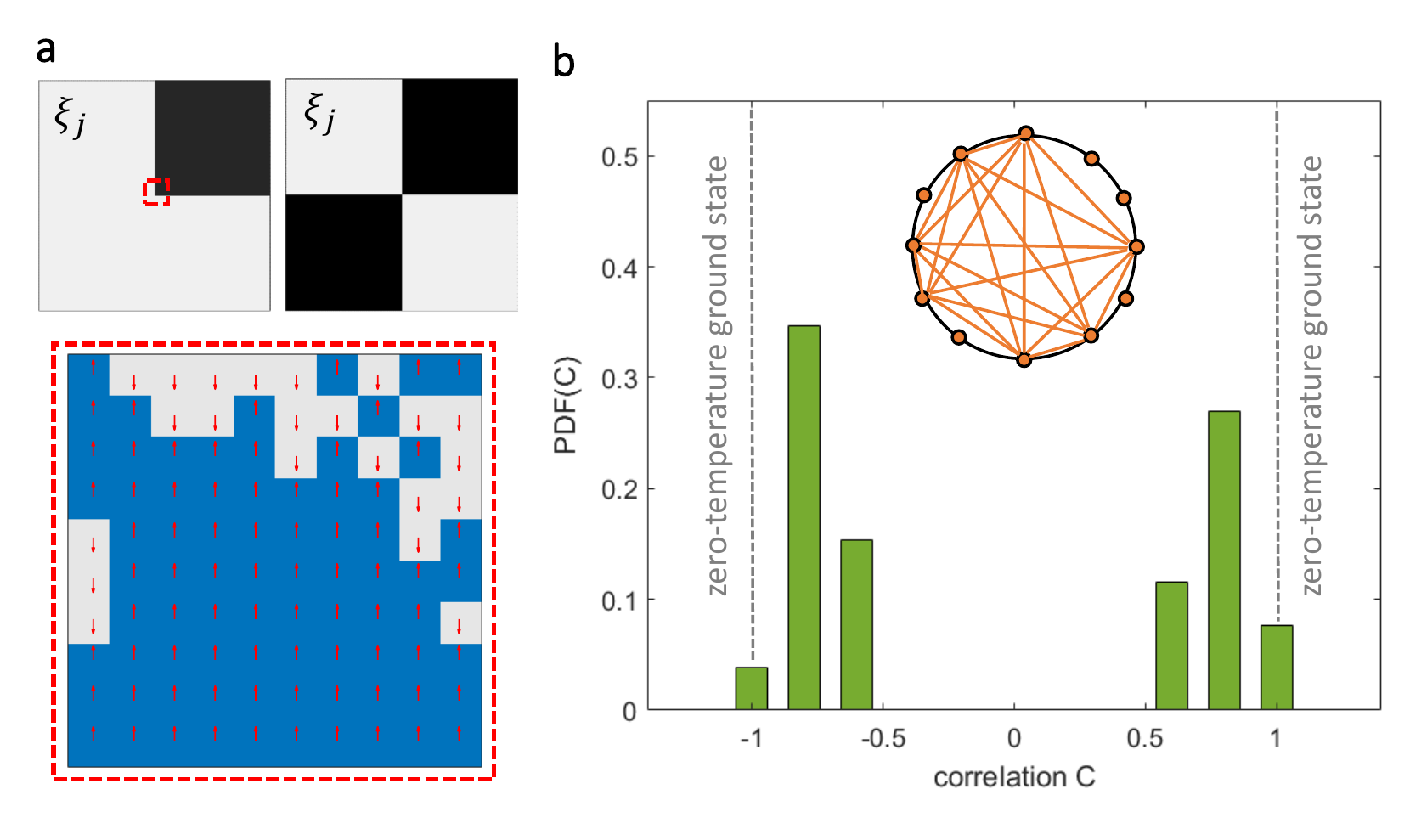} 
\vspace*{-0.5cm}
\caption{Programming the spin interaction by amplitude modulation. 
(a) Examples of coupling configurations ($N=10^4$, top panels) made of random blocks in which the interaction assumes two positive 
values $(\xi_j=0,\xi_0>0)$. The corresponding spin ground state observed in the red box region is shown in the bottom panel.
(b) Measured probability distribution of the correlation $C$ between the ground state and the couplings for the Mattis models. The inset
shows a corresponding M\"obius-Ladder graph with connected and unconnected nodes. 
}
\vspace{-0.1cm}
\label{Figure4}
\end{figure}

We investigate other Ising models, by tailoring the spin couplings. As suggested by Eq.~(\ref{eq:interft}), Mattis spin-glasses can be realized varying the input amplitudes $\xi_i$ and keeping a point-like target image
($\tilde I_T[2W(i-j)]\simeq \text{const}$). 
For these experiments, the SLM is split into two independent parts \cite{Ritsch-Marte2008}. A portion of the SLM is used for amplitude modulation to generate controlled $\xi_i$ distributions, that are imaged pixel by pixel on the second portion, where binary phase modulation and spin dynamics occur.
We implement coupling matrices $J_{jh}\propto \xi_j \xi_h$ made of large random blocks with strongly ($\xi_j=\xi_0$) and weakly ($\xi_j=0$) interacting spins [Fig.~\ref{Figure4}(a)]. 
Following the theoretical solution of the Mattis model \cite{Nishimori2001},
the expected spin ground state is identical to the interaction configuration $\xi_j$, or to its reversal, except for the weakly interacting regions where spins are randomly oriented. In our photonic simulation,  
we quantify the fidelity of the measured inhomogeneous ferromagnetic ground state by the spatial correlation $C=\sum_{j} \sigma_{j} \xi_j/\xi_0$. 
$C=\pm 1$ for the ideal Mattis model in the lowest energy state.
Figure~\ref{Figure4}(b) shows that the measured ground states are strongly correlated or anticorrelated with the interaction matrix, as expected.  Since in the Mattis models a minimal amount of noise introduces frustration \cite{Nishimori2001}, the differences between the machine solutions and the ideal ones are due to the non-zero effective temperature of the system. 

In conclusion, we have demonstrated that spatial light modulation can be exploited to find the ground state of Ising Hamiltonians. 
By using binary phases on the wavefront of an amplitude modulated laser beam and a detection and feedback method, 
we optically calculate the low-energy ferromagnetic spin configuration. The ground states display finite-size scaling effects and
mean-field properties at a fixed temperature. This finding opens the way to photonic simulations of phase-transition phenomena.
The platform naturally hosts tens of thousands of spins (not limited to binary spins, when adopting multilevel phase modulations)
and is scalable to larger sizes. The speed of our machine is limited only by the SLM response, camera rate, and data processing. 
The iteration time can be potentially reduced to few milliseconds with the most recent technologies \cite{Liu2017}. Moreover, a recent theoretical proposal for optical circuits \cite{RoquesCarmes2018} suggests a possible direction for further reducing the steps performed digitally using wave-mixing devices.
Our method, employing fast and low-loss optical computation, may also find application in alleviating energy-consuming operations in electronics,
as large matrix multiplications and Fourier transforms.
Our approach can be extended to light pulses modulated in space and time, even including the quantum optical regimes in which the coherent laser source is replaced by non-classical light.
Similar large-scale simulators may also be conceived with quantum wavepackets as in ultracold gases, and Bose-Einstein condensates,
by proper control and preparation of the initial states.

\vspace*{0.2cm}
We acknowledge funding from Sapienza Ateneo, QuantERA ERA-NET Co-fund (Grant No. 731473, project QUOMPLEX), PRIN NEMO 2015, PRIN PELM 2017
and H2020 PhoQus project (Grant No. 820392).

\end{document}